\pdfoutput=1
\documentclass[%
aip,
amsmath,amssymb,
reprint,%
]{revtex4-1}

\usepackage{graphicx}
\usepackage{dcolumn}
\usepackage{bm}

\usepackage[utf8]{inputenc}
\usepackage[T1]{fontenc}
\usepackage{mathptmx}
\usepackage{etoolbox}

\usepackage{algpseudocode}
\usepackage{booktabs}   

\usepackage{hyperref}
\hypersetup{
	colorlinks=true,
	linkcolor=blue,
	citecolor=blue,
	urlcolor=blue
}
\newcounter{myalg}
\renewcommand{\themyalg}{\arabic{myalg}}

\makeatletter
\def\@email#1#2{%
	\endgroup
	\patchcmd{\titleblock@produce}
	{\frontmatter@RRAPformat}
	{\frontmatter@RRAPformat{\produce@RRAP{*#1\href{mailto:#2}{#2}}}\frontmatter@RRAPformat}
	{}{}
}%
\makeatother
\begin{document}
	
	\preprint{AIP/123-QED}
	
	\title[Spatially Masked Regression]{Spatially Masked Regression Reveals Local and Distributed Predictability in Electrophysiological Recordings}
	\author{Maryam Ostadsharif Memar}
	\affiliation{ 
		Department of Electrical and Computer Engineering, IUT.
	}%
	
	\author{Nima Dehghani}
	\email{nima.dehghani@mit.edu.}
	\affiliation{%
		McGovern Institute for Brain Research, Massachusetts Institute of Technology (MIT).
	}%
	
	\date{\today}
	
	\keywords{functional connectivity, spatial redundancy, distributed predictability, neural signal reconstruction, local masking, distance correlation, volume conduction, cross-subject generalization}
	
	\begin{abstract}
		Neural recordings are often interpreted as local measurements, yet the signal at any one sensor can also reflect structured activity distributed across the broader network. This raises a basic question: to what extent does an electrode's signal reflect local versus distributed information in the underlying system? More specifically, how much of an electrode's activity is carried by its immediate neighborhood, and how much is embedded more broadly across the array? We address this with a Spatially Masked Regression (SMR) framework that reconstructs each electrode's timeseries from the remaining electrodes while excluding a configurable neighborhood around the target. By progressively increasing this mask, spatial locality becomes an experimental control for quantifying how much predictive information survives after nearby channels are withheld. We apply SMR to intracranial EEG with heterogeneous electrode coverage and to scalp EEG with standardized montages over sensorimotor cortex. Using distance correlation between original and reconstructed signals, we find strong within-subject reconstruction in both modalities, substantial residual predictability even when local neighbors are excluded, and markedly stronger cross-subject transfer in EEG than in iEEG. Masking shows that nearby electrodes contribute strongly to reconstruction but do not account for all of it, indicating that individual channels reflect both local redundancy and broader distributed structure. Surrogates that preserve selected marginal or spectral properties while disrupting phase structure or temporal ordering substantially reduce performance, supporting the conclusion that SMR depends on structured temporal and cross-channel organization rather than on marginal statistics alone. These results position SMR as an interpretable framework for quantifying the balance between local and distributed information in recordings.
		\footnote{Context $\&$ overview:\\ \url{https://neurovium.science/posts/pblog-SMR-local-global}
			\\ Code $\&$ experiments: \\ \url{https://github.com/neurovium/SpatiallyMaskedRegression}}
	\end{abstract}
	
	\maketitle
	

	\section{Introduction}
	Neural activity is organized across multiple spatial scales: from synapses and microcircuits to distributed networks that coordinate distant cortical and subcortical regions \cite{Buzsaki2004Oscillations,Bullmore2009Complex}. Electrophysiological recordings such as electroencephalography (EEG) and intracranial EEG (iEEG) occupy a particularly revealing position in this hierarchy. These modalities play a crucial role in both research and clinical investigations of functional brain dynamics \cite{Parvizi2018, Mercier2022, Zhang2023}. They are often described as ``local field'' measurements, yet many of their most robust and interpretable signatures---sensorimotor rhythms, large-scale synchrony, task-locked modulations---are difficult to attribute to strictly local generators \cite{Buzsaki2004Oscillations}. The same sensor can simultaneously inherit strong local structure from the immediate neighborhood and express global structure imposed by distributed network dynamics. This duality motivates a basic question that is surprisingly hard to operationalize: to what extent does an electrode’s signal reflect \emph{local} activity versus \emph{global} structure that is distributed across the network?
	
	A standard response is to frame the question in terms of interregional interaction. Rather than treating each region in isolation, modern systems neuroscience emphasizes the pattern of dependencies among regions, often summarized as functional connectivity---statistical dependence between signals measured at different locations \cite{Friston2011ConnectivityReview,Lang2012BrainConnectivitySurvey}. In practice, these dependencies are frequently estimated from electrophysiology, where temporal resolution is high and where oscillatory coordination offers natural candidate mechanisms for communication \cite{Srinivasan2007ScalesCoherence}. A broad toolbox exists to examine such relationships: Pearson correlation, coherence, and phase-based measures such as phase-locking value; information-theoretic measures such as mutual information and transfer entropy; and network/graph summaries built on those pairwise relations \cite{Lachaux1999PLV, Varela2001PhaseSynchronization,Ince2017GCMI,Vicente2011TransferEntropy,Bullmore2009Complex}. These approaches have been indispensable for describing mesoscale and macroscale organization, but they also tend to blur an important distinction between \emph{local redundancy} and \emph{distributed predictability}.
	
	The reason is partly biophysical. Field potentials measured on the scalp or on cortical surfaces are shaped by spatial mixing and by the conductive properties of tissue, which induce smoothness and shared variance across nearby sensors \cite{Nolte2004ImagCoherency,Srinivasan2007ScalesCoherence}. At finer scales, extracellular filtering produces strong distance- and frequency-dependent attenuation, causing fast components to decay more steeply with distance than slower components, thereby reinforcing short-range similarity in many regimes \cite{Bedard2004FreqFiltering}. At larger scales, macroscopic tissue properties can imprint broadband structure on field recordings \cite{Bedard2009MacroscopicLFP1f}, and simultaneous EEG/MEG spectral scaling results suggest that modeling assumptions about the extracellular medium can matter for how we interpret these signals \cite{Dehghani2010EEGMEGScaling}. Consequently, pairwise connectivity estimates at the sensor level can be dominated by local, almost redundant dependencies; in that setting, strong local coupling can mask weaker but functionally meaningful long-range structure \cite{Nolte2004ImagCoherency,Schoffelen2009SourceConnectivity}.
	
	Network-level summaries inherit related limitations. Graph-theoretic analyses provide an appealing language for large-scale organization \cite{Bullmore2009Complex}, but the typical construction starts from symmetric association measures and then treats edges as interchangeable objects, which can obscure the directionality and the \emph{reconstructability} of activity at one site from activity elsewhere \cite{Hallquist2019BraveNewSmallWorld}. More broadly, many popular connectivity measures answer ``are these two signals related?'' rather than ``how much of this signal is contained in the rest of the array, and where is that information coming from?'' The latter is an operational question about representation and redundancy in multichannel recordings, and it becomes especially salient when sensor geometry and coverage vary across individuals, as in clinical iEEG.
	
	Here we take a complementary, reconstruction-based perspective. Instead of using pairwise connectivity as the primitive, we ask: \emph{how well can the signal at a given electrode be predicted from the rest of the array?} This viewpoint resonates with a growing line of work that treats neural recordings as partially observed samples of a structured dynamical system and leverages that structure to infer missing or unobserved activity. Examples span Gaussian-process frameworks for inferring activity at unobserved locations from sparse intracranial recordings \cite{Owen2020GaussianProcessECoG}, latent dynamical models that denoise and reconstruct single-trial population activity (albeit for the spiking data) \cite{Pandarinath2018LFADS}, and sparse multivariate autoregressive models that estimate directed interactions in high-dimensional time series \cite{ValdesSosa2005SparseMVAR}. Our goal is not to replace these frameworks, but to use reconstruction as a diagnostic: a way to \emph{measure} how information about one channel is distributed across space.
	
	To do so, we introduce a Spatially Masked Regression (SMR) model that explicitly separates local from nonlocal contributions. For each target electrode, SMR learns a linear mapping from all other electrodes, but imposes a spatial mask that excludes a predefined neighborhood around the target. By increasing the size (or strength) of this mask, we remove local predictors in a controlled way and force the model to rely on progressively more distant signals. This turns spatial locality into an experimental knob: at zero masking, the model reflects the familiar locality-dominated regime; at strict masking, it probes how much of the target signal is embedded in distributed structure beyond the immediate neighborhood.
	
	We evaluate the local--global trade-off across two recording modalities that sit at different points on the spatial mixing spectrum. Scalp EEG uses standardized montages that support direct sensor-level comparison across subjects \cite{Jurcak2007EEGMontage} and is expected to emphasize more global structure due to spatial smoothing and volume conduction \cite{Srinivasan2007ScalesCoherence,Nolte2004ImagCoherency}. In contrast, iEEG provides high signal quality and spatial specificity, but electrode placement is clinically constrained and heterogeneous across participants, complicating cross-subject comparisons at the sensor level \cite{Peterson2022AJILE12}. By applying the same SMR framework to a long-term naturalistic iEEG dataset (AJILE12) \cite{Peterson2022AJILE12} and to an EEG dataset with uniform sensor coverage over sensorimotor regions \cite{Jurcak2007EEGMontage}, and by comparing within-subject and cross-subject performance, we ask: \emph{How distributed are the linear dependencies in these signals, and how consistent is that distribution across individuals and modalities?}
	
	Throughout, we quantify reconstruction quality using Distance Correlation, which provides a dependence measure sensitive to broad classes of relationships while remaining interpretable as a scalar score \cite{Szekely2007DistanceCorrelation}. This choice allows us to compare regimes where linear predictability is high but nonlinear dependence may also be present, and it supports a consistent evaluation of masking effects across datasets and recording geometries.

	\begin{figure*}[t]
		\begin{center}
			\includegraphics[width=\textwidth]{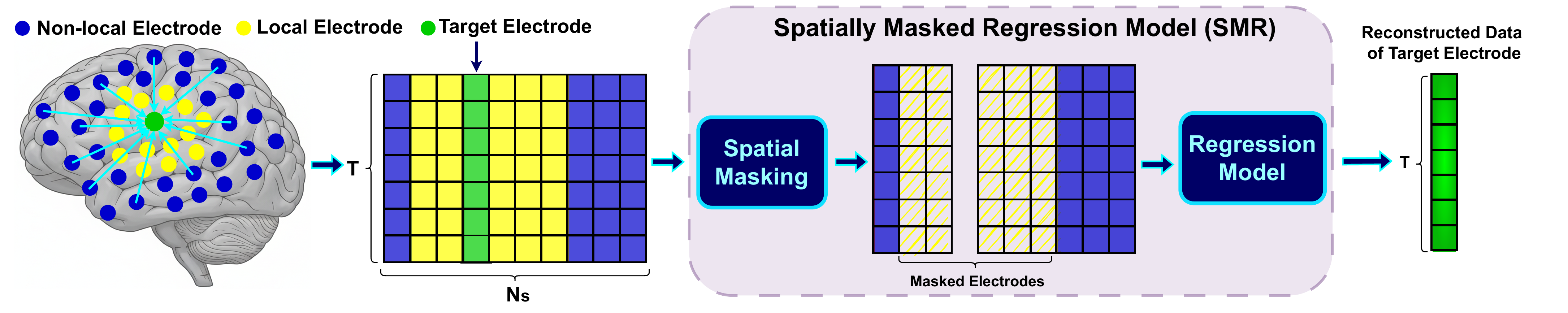}
		\end{center}
		\vspace{-5pt}
		\caption{Spatially Masked Regression (SMR) framework. For each target electrode, SMR reconstructs the target time series from the remaining electrodes while excluding a predefined local neighborhood through a binary spatial mask. The masking step suppresses nearby predictors, and the regression step estimates the target from the remaining nonlocal channels. Varying the mask size provides a controlled way to compare local and distributed contributions to reconstruction. Here, $T$ denotes the number of time samples and $N_s$ the number of electrodes for subject $s$.}     
		\label{fig1}
	\end{figure*}

	\section{Material and Methods}
	\subsection{Dataset}
	
	\paragraph{AJILE12 iEEG} \cite{Peterson2022AJILE12}: This dataset comprises iEEG recordings collected from 12 subjects (8 males, 4 females; mean age 29 ± 7 years) during clinical epilepsy monitoring. Electrode implantation was limited to a single hemisphere per subject (5 right, 7 left), and electrode locations varied significantly across subjects. Fig.~\ref{fig_app} shows the electrode configurations and the distribution of implanted brain regions across subjects. During this period, subjects engaged in routine hospital activities such as eating, watching TV, and socializing. Simultaneously, video recordings were used to identify instances of upper-limb movement, which served as behavioral annotations aligned with the neural data. The neural data were preprocessed (DC removal and linear detrending, 
	0.5--150~Hz Butterworth band-pass, 60~Hz notch filtering, a 120~Hz low-pass for 
	residual high-frequency noise, \(|z| > 5\) artifact rejection by per-channel 
	replacement with the channel median, Common Average Reference, and per-channel 
	z-scoring across time), and were segmented into 2-second trials centered on 
	movement events. 
	
	\paragraph{Upper Limb Movements EEG} \cite{Ofner2017eeg}: This EEG dataset was recorded to investigate both motor execution and motor imagery of upper-limb movements. The dataset comprises recordings from 15 healthy subjects (9 female; aged 22–40 years), each performing six distinct right-arm movements, including elbow flexion/extension, forearm supination/pronation, and hand opening/closing, along with a rest condition. Each subject completed two sessions on different days: one involving actual movement execution and the other involving kinesthetic motor imagery. EEG signals were acquired from 61 scalp electrodes covering frontal, central, parietal, and temporal regions. After acquisition, EEG signals were band-pass filtered, notch-filtered at line frequency, and z-scored per channel before being segmented into trials.
	
	\begin{table}[ht]
		\centering
		\caption{Notations used in the paper}
		\begin{ruledtabular}
			\begin{tabular}{ll}
				\textbf{Symbol} & \textbf{Description} \\ \hline
				$s$ & Target subject \\
				$s'$ & Reference subject \\
				$\mathcal{N}^{(s)}(i)$ & Neighborhood of electrode $i$ for subject $s$ \\
				$k$ & Number of nearest electrodes \\
				$N_s$ & Number of channels of subject $s$ \\
				$K_{ij}$ & Binary spatial mask between electrodes $i$ and $j$ \\
				$w_{ij}$ & Learnable weight from electrode $j$ to electrode $i$ \\
				$T$ & Number of time points per channel \\
				$M_{s'}^{\mathrm{learn}}$ & Learned relationship matrix of subject $s'$ \\
				$f_s(\cdot,\cdot)$ & Subject-specific mapping function \\
				$M_s^{\mathrm{est}}$ & Estimated relationship matrix of subject $s$ \\
				$C_{ss'}(i,j)$ & Pearson correlation between electrodes $i$ and $j$ \\
				$\mathcal{T}$ & Set of temporal lags \\
				DistCorr & Distance correlation \\
			\end{tabular}
		\end{ruledtabular}
	\end{table}
	
	\subsection{Spatially Masked Regression Model}
	\label{model_arch}
	To probe how much information about a target electrode is available locally versus elsewhere in the array, we use a \textit{Spatially Masked Regression (SMR)} model. Figure~\ref{fig1} summarizes the framework. For each target channel, SMR reconstructs the target signal from the remaining channels while explicitly excluding a predefined local neighborhood.
	
	The motivation is straightforward. In multichannel electrophysiology, nearby electrodes often carry highly similar signals because brain activity varies smoothly in space and because field recordings are spatially mixed due to volume conduction. If all predictors are allowed, a linear model will typically assign large weights to nearby channels and comparatively small weights to more distant ones. Such a model is useful for reconstruction, but it is poorly suited for asking how much predictive structure remains once the dominant local contribution is removed thus limiting its effectiveness for uncovering broader spatial dependencies..
	
	We therefore define, for each electrode $i$ of subject $s$, a neighborhood set $\mathcal N^{(s)}(i)$ containing the channels spatially closest to the target. For both modalities, $\mathcal N^{(s)}(i)$ is the set of $k=9$ electrodes whose recording sites are nearest to electrode $i$ under a fixed projected Euclidean distance: for EEG we use the in-plane distance in the $(x,y)$ coordinates of the standardized scalp montage, while for iEEG we use the $(x,z)$ projection of the MNI electrode coordinates (the electrode itself, at distance zero, is excluded). This coordinate-based construction yields a neighborhood of fixed size across electrodes and avoids relying on either a standardized montage labeling or on subject-specific atlas-based parcellation, which is necessary for the AJILE12 cohort because electrode placement is heterogeneous across patients \cite{Peterson2022AJILE12}. The corresponding binary mask $K_{ij}^{(s)} \in \{0,1\}$ is
	
	\[
	K_{ij}^{(s)} = 
	\begin{cases}
		1, & j \in \mathcal{N}^{(s)}(i),\\
		0, & \text{otherwise.}
	\end{cases}
	\]
	
	When $K_{ij}^{(s)}=1$, channel $j$ is excluded from the reconstruction of channel $i$. The predicted (reconstructed) signal $\widehat{x}_i^{(s)}$ for target electrode $i$ is therefore
	
	\begin{equation}
		\label{eq:smr_prediction}
		\widehat{x}_i^{(s)} = \sum_{j=1}^{N_s} (1 - K_{ij}^{(s)}) \, w_{ij}^{(s)} \, x_j^{(s)},
	\end{equation}

	where $x_j^{(s)}$ is the signal recorded from electrode $j$, $w_{ij}^{(s)}$ is the learned weight assigned to that predictor, and $N_s$ is the number of electrodes for subject $s$. The learned weights are asymmetric in general: $w_{ij}^{(s)}$ quantifies the contribution of channel $j$ to reconstructing channel $i$, not vice versa.
	
	This construction turns locality into an explicit experimental control. With weak or no masking, reconstruction is dominated by nearby channels. With strict masking, the model can rely only on nonlocal predictors. Comparing performance across these regimes provides an operational measure of how strongly each channel is represented in its immediate neighborhood versus more distributed activity across the array.
	
	\subsection{Model Training}
	For each target electrode, SMR learns a separate set of regression weights by minimizing a regularized reconstruction loss using mini-batch stochastic optimization with Adam \cite{Kingma2015adam}. The data term is the mean absolute error (L1 loss) between the original and reconstructed signals. Early stopping based on the validation loss was used to limit overfitting, and Elastic Net regularization \cite{Zou2005elasticnet} was included to encourage a sparse but not overly concentrated weight profile. Specifically, the L1 component promotes sparsity, whereas the L2 component discourages excessively large individual weights and favors a more distributed set of coefficients across predictors.
	
	Given the original signal \(x_i^{(s)}(t)\) and the reconstructed signal \(\widehat{x}_i^{(s)}(t)\) for subject \(s\), electrode \(i\), and time index \(t=1,\ldots,T\), the optimization problem is
	
	\begin{align}
		\label{eq:reconstruction_error_instant}
		\min_{w_i^{(s)}} \;
		& \frac{1}{T} \sum_{t=1}^{T}
		\Bigg|
		x_i^{(s)}(t)
		-
		\sum_{j=1}^{N_s} (1-K_{ij}^{(s)})\, w_{ij}^{(s)}\, x_j^{(s)}(t)
		\Bigg| \nonumber \\
		& + \lambda_1 \lVert \theta_i^{(s)} \rVert_1 + \lambda_2 \lVert \theta_i^{(s)} \rVert_2^2,
		\qquad \text{s.t. } w_{ij}^{(s)}=0 \; \forall j \in \mathcal{N}^{(s)}(i),
	\end{align}
	
	Here, \(\theta_{i}^{(s)} = \big(w_{i}^{(s)},\, b_{i}^{(s)}\big)\) denotes all 
	trainable parameters used to reconstruct channel \(i\) (the linear-layer weight 
	vector \(w_{i}^{(s)}\) and the scalar bias \(b_{i}^{(s)}\)). The first term 
	penalizes reconstruction error, while the L1 and L2 penalties, scaled by 
	\(\lambda_1\) and \(\lambda_2\), are applied to all trainable parameters 
	\(\theta_{i}^{(s)}\) — including the bias — and control sparsity and coefficient 
	magnitude, respectively. In practice, the spatial mask and the explicit zero 
	constraint are equivalent ways of enforcing local exclusion; both are written 
	here for clarity.
	
	\subsection{Model Evaluation}
	\label{sec_model_eval}
	
	We evaluated the performance of our model under two experimental paradigms: \textbf{intra-subject} and \textbf{cross-subject} settings. These settings were designed to assess the model’s ability to learn subject-specific neural patterns and its capacity to generalize across individuals. 
	
	\subsubsection{Intra-subject evaluation}
	
	\stepcounter{myalg}
	\makeatletter\def\@currentlabel{\themyalg}\makeatother\label{alg:intra} 
	\noindent\hrule
	\vspace{1mm}
	\noindent \textbf{Algorithm \themyalg:} Intra-Subject Evaluation of the SMR Model
	\vspace{1mm}
	\hrule
	\vspace{1mm}
	\begin{algorithmic}[1]
		\Require Signals $X^{(s)} \in \mathbb{R}^{N_s \times T}$, neighborhood mask $K^{(s)} \in \{0,1\}^{N_s \times N_s}$, regularization $\lambda_1, \lambda_2$
		\Ensure Learned relationship matrix $M_{s}^{\mathrm{learn}}$
		\Statex \hrulefill
		
		\For{$i = 1$ to $N_s$}
		\State Initialize $w_i^{(s)}$ randomly
		\State $\widehat{x}_i^{(s)} \gets 0$
		\Repeat
		\For{$j = 1$ to $N_s$}
		\If{$K_{ij}^{(s)} = 0$}
		\State $\widehat{x}_i^{(s)} \gets \widehat{x}_i^{(s)} + w_{ij}^{(s)} x_j^{(s)}$
		\EndIf
		\EndFor
		\State Compute loss: 
		\Statex \hspace{1em} $\mathcal{L}^{(s)} = \frac{1}{T}\sum_{t} \big| x_i^{(s)}(t) - \widehat{x}_i^{(s)}(t)\big| + \lambda_1 \| \theta_i^{(s)} \|_1 + \lambda_2 \| \theta_i^{(s)} \|_2^2$
		\State Update $w_i^{(s)}$ using Adam optimizer
		\Until{convergence or early stopping}
		\State $M_{s}^{\mathrm{learn}}[i, :] \gets w_i^{(s)}$
		\EndFor
		\State \textbf{return} $M_{s}^{\mathrm{learn}}$
	\end{algorithmic}
	\hrule
	\vspace{2mm}
	
			
	
	In the intra-subject setting, all training and evaluation data came from the same subject. For each subject, the full dataset was first z-scored across trials and then partitioned into training, validation, and test subsets in a 
	64/16/20 \% split, obtained by two nested 80/20 random splits (random seed fixed). The training set was used for gradient updates, the validation set for early stopping, and the test set for reporting reconstruction quality. Training was terminated with patience-based monitoring of the epoch-averaged validation loss, and the checkpoint achieving the lowest validation loss was retained for subsequent evaluation on the held-out test set.
	
	For each target electrode \(i\), a separate reconstruction model was trained using only the unmasked channels as predictors. 
	This procedure yielded a subject-specific asymmetric relationship matrix \(M_{s}^{\mathrm{learn}} \in \mathbb{R}^{N_s \times N_s}\), where \(M_{s}^{\mathrm{learn}}[i,j]\) denotes the learned contribution of electrode \(j\) to reconstructing electrode \(i\) for subject \(s\). Because the reconstruction mapping is directional, \(M_{s}^{\mathrm{learn}}[i,j]\) and \(M_{s}^{\mathrm{learn}}[j,i]\) need not be equal. Reconstruction quality was then quantified on the held-out test data using distance correlation. Algorithm~\ref{alg:intra} summarizes the intra-subject evaluation procedure. These learned matrices were subsequently used in the cross-subject analyses.
	Model parameters were optimized with Adam using an L1 reconstruction loss together with L1 and L2 penalties on the trainable parameters.
	
	\subsubsection{Cross-subject evaluation}
	\label{cross_sub_eval}
	
	\stepcounter{myalg}
	\makeatletter\def\@currentlabel{\themyalg}\makeatother\label{alg:cross} 
	\noindent\hrule
	\vspace{1mm}
	\noindent \textbf{Algorithm \themyalg:} Cross-Subject Evaluation of the SMR Model (per-electrode donor selection)
	\vspace{1mm}
	\hrule
	\vspace{1mm}
	\begin{algorithmic}[1]
		\Require Target subject signals $X^{(s)}$, reference subjects $\{X^{(s')}\}_{s' \neq s}$, reference matrices $\{M_{s'}^{\mathrm{learn}}\}_{s' \neq s}$
		\Ensure  Row-wise composed transferred matrix $M_s^{\mathrm{est}}$
		\Statex \hrulefill
		
		\State Split $X^{(s)}$ into validation $X^{(s)}_{\mathrm{val}}$ and test $X^{(s)}_{\mathrm{test}}$ (50/50)
		\For{each reference subject $s' \neq s$}
		\State Compute the cross-subject correlation matrix $\mathbf C_{ss'} \in \mathbb{R}^{N_s \times N_{s'}}$
		\State Solve the rectangular assignment problem
		\[
		\mathbf P^{(s \leftarrow s')} =
		\arg\max_{\mathbf P \in \mathcal{P}}
		\sum_{r,c} P[r,c]\,C_{ss'}[r,c],
		\]
		where $\mathcal{P}=\{\mathbf P:\mathbf P\mathbf 1 \le \mathbf 1,\ \mathbf P^\top \mathbf 1 \le \mathbf 1\}$, solved by the Hungarian algorithm
		\State Transfer the reference matrix:
		\[
		M_{s \leftarrow s'}^{\mathrm{est}}
		\gets
		\mathbf P^{(s \leftarrow s')} \, M_{s'}^{\mathrm{learn}} \, \mathbf P^{(s \leftarrow s')\top}
		\]
		\EndFor
		\For{$i = 1$ to $N_s$}
		\State $s'_i \gets \arg\max_{s' \neq s} \mathrm{DistCorr}\!\big(\overline{x}_{i,\mathrm{val}}^{(s)},\ M_{s \leftarrow s'}^{\mathrm{est}}[i,:]\,\overline{X}_{\mathrm{val}}^{(s)}\big)$
		\State $M_s^{\mathrm{est}}[i,:] \gets M_{s \leftarrow s'_i}^{\mathrm{est}}[i,:]$
		\EndFor
		\State \textbf{return} $M_s^{\mathrm{est}}$
		\Statex \hfill\(\triangleright\) where \(\overline{X}_{\mathrm{val}}^{(s)}\) denotes the trial-averaged signals on the validation subset and \(\overline{x}_{i,\mathrm{val}}^{(s)}\) its \(i\)-th row.
	\end{algorithmic}
	\hrule
	\vspace{2mm}

	In the cross-subject setting, one subject was treated as the target subject \(s\), and relationship matrices learned from the remaining subjects \(s' \neq s\) were transferred to the target. The goal of this analysis was not to fit target-specific weights, but to test how well inter-electrode structure learned in one subject could be transferred to another. Here, \(M_{s'}^{\mathrm{learn}}\) denotes a matrix learned directly from a reference subject \(s'\), whereas \(M_{s \leftarrow s'}^{\mathrm{est}}\) denotes the corresponding matrix after transfer into the electrode space of target subject \(s\). After per-electrode selection among candidate reference subjects (see below), 
	we denote the resulting row-wise composed matrix used for evaluation by 
	\(M^{\mathrm{est}}_{s}\). By construction, different rows of 
	\(M^{\mathrm{est}}_{s}\) may originate from different reference subjects.

	For EEG, electrode positions are standardized across subjects, so the learned matrix from a reference subject can be transferred directly to the target subject. In iEEG, the number and anatomical placement of electrodes differ substantially across patients, making direct transfer infeasible. To enable transfer in that setting, we used a \textit{correlation-based electrode mapping} (CBEM) procedure as an alignment step prior to matrix transfer. The resulting transferred matrices were computed for all candidate reference subjects, and the matrix used for evaluation was selected using a held-out validation subset from the target subject. This step selected among transferred matrices only and did not involve fitting target-specific weights.
	
	For each target-reference pair \((s,s')\), we computed a cross-subject correlation matrix \(C_{ss'} \in \mathbb{R}^{N_s \times N_{s'}}\), where \(N_s\) and \(N_{s'}\) denote the numbers of electrodes in the target and reference subjects, respectively. Its entries are obtained by averaging Pearson correlations across the set of 
	paired trials \(R_{ss'} = \{1,\ldots,\min(R_s, R_{s'})\}\) shared between 
	subjects \(s\) and \(s'\), where \(\rho(\cdot,\cdot)\) denotes the Pearson 
	correlation between two equal-length time series, and 
	\(x_{i}^{(s),r}\) is the time series of electrode \(i\) on trial \(r\) of 
	subject \(s\). Trials yielding NaN correlations (e.g., from constant segments) 
	are excluded from the mean.
	
	\begin{equation}
		C_{ss'}(i,j)
		=
		\frac{1}{|\mathcal{R}_{ss'}|}
		\sum_{r \in \mathcal{R}_{ss'}}
		\rho\!\left(x_i^{(s),r},\, x_j^{(s'),r}\right),
	\end{equation}
	
	\begin{figure}[t]
		\begin{center}
			\includegraphics[width=\columnwidth]{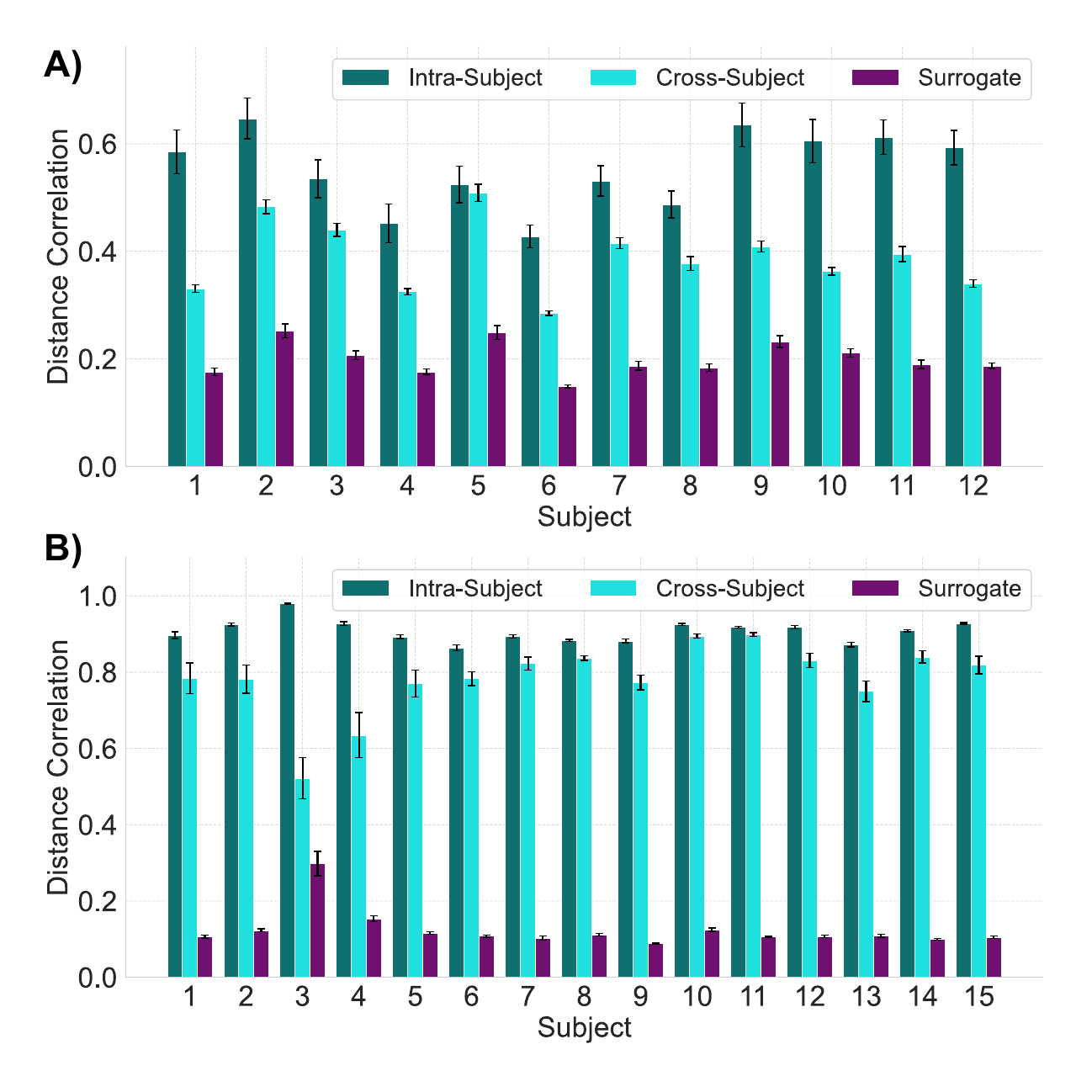}
		\end{center}
		\caption{Subject-wise reconstruction performance in the intra-subject and cross-subject settings for (A) iEEG and (B) EEG. For each subject, bars show the mean distance correlation (DistCorr) across electrodes.}
		\label{fig2}
	\end{figure}
	
	To obtain a one-to-one correspondence between the electrode sets of two subjects, we formulated the alignment as a rectangular assignment problem. This provides an assignment-based view of correlation-guided electrode matching and is closely related to Hungarian-style maximum-weight bipartite matching \cite{Munkers1957hungarian}. Specifically, we sought a binary matrix \(\mathbf P^{(s\leftarrow s')} \in \{0,1\}^{N_s \times N_{s'}}\) that maximizes the total matched correlation,
	
	\begin{equation}
		\mathbf P^{(s\leftarrow s')}
		=
		\arg\max_{\mathbf P \in \mathcal{P}}
		\sum_{r,c} P[r,c]\,C_{ss'}[r,c],
	\end{equation}
	where
	\[
	\mathcal{P}
	=
	\left\{
	\mathbf P:
	\mathbf P\mathbf 1 \le \mathbf 1,\;
	\mathbf P^\top \mathbf 1 \le \mathbf 1
	\right\}
	\]
	enforces that each electrode is used at most once. 
	
	Given this mapping, the learned matrix from the reference subject was transferred into the target electrode space according to
	\begin{equation}
		M_{s \leftarrow s'}^{\mathrm{est}}
		=
		\mathbf P^{(s\leftarrow s')} \, M_{s'}^{\mathrm{learn}} \, \mathbf P^{(s\leftarrow s')\top}.
	\end{equation}
	
	For iEEG, the procedure was repeated for all candidate reference subjects 
	\(s' \neq s\), yielding one transferred matrix \(M^{\mathrm{est}}_{s \leftarrow s'}\) 
	per donor. Donor selection was then performed independently for each target 
	electrode \(i\): the target subject's trials were split 50/50 into a validation 
	and a test subset, and the donor \(s'\) whose reconstruction of the 
	trial-averaged signal at electrode \(i\) on the validation subset attained the 
	highest DistCorr was chosen for that electrode. The corresponding row of 
	\(M^{\mathrm{est}}_{s \leftarrow s'}\) was then used to evaluate electrode \(i\) on 
	the held-out test subset. The resulting \(M^{\mathrm{est}}_{s}\) is therefore a 
	row-wise composition: each row is drawn from the transferred matrix of the donor 
	selected for that electrode. This selection step operated only over transferred 
	matrices; no target-specific weights were fit during cross-subject evaluation. 
	
	Unlike the intra-subject evaluation, which reports DistCorr as the mean across 
	single-trial reconstructions, cross-subject DistCorr for iEEG is computed on the 
	trial-averaged signals: both the original target trace 
	\(\overline{x}_{i}^{(s)}\) and the reconstructed trace 
	\(\overline{\hat{x}}_{i}^{(s)} = M_{s}^{\mathrm{est}}[i,:]\,\overline{X}^{(s)}\) 
	are obtained by averaging across the trials in the corresponding split 
	(validation or test) and individually z-scored prior to computing DistCorr.
	The full procedure is summarized in Algorithm~\ref{alg:cross}.
	
	\subsection{Lagged Variation of the Spatially Masked Regression Model}
	The baseline SMR model is instantaneous: it reconstructs each target channel at time \(t\) from the other channels at the same time point. To test whether short temporal offsets improve reconstruction, we also considered a lagged variant in which the predictors include delayed versions of electric potential at the other channels.
	
	For target electrode \(i\), the lagged model is
	
	\begin{equation}
		\widehat{x}^{(s)}_i(t) = 
		\sum_{\tau \in \mathcal{T}} \sum_{j = 1}^{N_s}
		(1-K^{(s)}_{ij}) \, w^{(s)}_{ij}[\tau] \, x^{(s)}_j(t-\tau),
	\end{equation}
	
	where \(\mathcal{T}\) denotes the set of temporal lags and \(K^{(s)}_{ij}\) is the same spatial mask used in the instantaneous model. The coefficients \(w^{(s)}_{ij}[\tau]\) quantify time-lagged predictive contributions at delay \(\tau\). In the implementation used here, the lag set was fixed to 
	\(T = \{0, 20, 30, 50, 60\}\,\mathrm{ms}\), with the \(0\) ms lag corresponding 
	to the instantaneous branch and the remaining lags each associated with an 
	independent set of learnable weights. Each lag was converted to a sample offset 
	\(\lceil (\tau/1000)\,f_s \rceil\) using the sampling frequency \(f_s\). The corresponding regularized objective is
	
	\begin{multline}
		\label{eq:reconstruction_error_lagged}
		\min_{w_i^{(s)}} \;
		\frac{1}{T} \sum_{t=1}^{T}
		\Bigg|
		x_i^{(s)}(t)
		-
		\sum_{\tau \in \mathcal{T}} \sum_{j=1}^{N_s}
		(1-K^{(s)}_{ij}) \, w^{(s)}_{ij}[\tau] \, x^{(s)}_j(t-\tau)
		\Bigg| \\
		+ \lambda_1 \lVert \theta_i^{(s)} \rVert_1 + \lambda_2 \lVert \theta_i^{(s)} \rVert_2^2,
		\qquad \text{s.t. } w_{ij}^{(s)}=0 \; \forall j \in \mathcal{N}^{(s)}(i).
	\end{multline}
	
	This lagged formulation is intended as an extension of the reconstruction framework, not as a causal model. It tests whether adding short temporal offsets yields additional predictive information beyond the instantaneous structure captured by the baseline SMR. In the codebase, the instantaneous and lagged variants are implemented as separate modules (\texttt{ReconModel} and \texttt{LaggedReconModel}); intra-subject runs for the Intra-Subject columns of Tables~\ref{tab:ieeg} and \ref{tab:eeg} use \texttt{ReconModel}, while the Lagged columns are obtained from a separate run in which \texttt{LaggedReconModel} is substituted at the model-construction step. Training settings (splits, early stopping, regularization, optimizer) are otherwise identical. The implementation is summarized in Algorithm~\ref{alg:lagged}.
	
	\vspace{3mm}
	
	\stepcounter{myalg}
	\makeatletter\def\@currentlabel{\themyalg}\makeatother\label{alg:lagged} 
	\noindent\hrule
	\vspace{1mm}
	\noindent \textbf{Algorithm \themyalg:} Lagged Variation of the SMR Model
	\vspace{1mm}
	\hrule
	\vspace{1mm}
	\begin{algorithmic}[1]
		\Require Signals $X^{(s)} \in \mathbb{R}^{N_s \times T}$, neighborhood mask $K^{(s)}$, lag set $\mathcal{T}$, regularization $\lambda_1,\lambda_2$
		\Ensure Learned lagged weight tensor $M_s^{\mathrm{learn}}$
		\Statex \hrulefill
		
		\For{$i = 1$ to $N_s$}
		\State Initialize lagged weights $w_{ij}^{(s)}[\tau]$.
		\Repeat
		\State Reconstruct $x_i^{(s)}(t)$ from all non-masked channels and all lags $\tau \in \mathcal{T}$.
		\State Compute the regularized lagged reconstruction loss.
		\State Update $w_{ij}^{(s)}[\tau]$ with Adam.
		\Until{convergence or early stopping}
		\State Store the learned lagged weights for target channel $i$.
		\EndFor
		\State \textbf{return} $M_s^{\mathrm{learn}}$
	\end{algorithmic}
	\hrule
	\vspace{1mm}
	
	
	\subsection{Hyperparameter Optimization}
	\label{sec:hyperparam}
	
	Initial hyperparameter values were chosen by a coarse search on the validation 
	set over learning rate, batch size, early-stopping patience, and the lag set 
	\(T\) for the lagged model. The regularization coefficients were fixed at 
	\(\lambda_1 = 10^{-5}\) and \(\lambda_2 = 10^{-4}\) for all reported runs. 
	Dropout was not used during training of the reported models. The final settings 
	were then fixed for evaluation on the held-out test data.
	
	\subsection{Evaluation Criteria: Distance Correlation Metric}
	To quantify the statistical dependence between the original and reconstructed signals, we used distance correlation (DistCorr), computed using the standard sample formulation based on doubly centered pairwise distance matrices \cite{Szekely2007DistanceCorrelation}. This measures statistical dependence between two variables and can capture both linear and nonlinear associations. For subject \(s\) and electrode \(i\), let $u = x_i^{(s)}, v = \widehat{x}_i^{(s)}$, denote the original and reconstructed signals, respectively, each represented as a length-\(T\) vector.
	
	We first define the pairwise distance matrices
	\begin{equation}
		a_{mn} = |u_m - u_n|, \qquad b_{mn} = |v_m - v_n|, \qquad m,n = 1,\ldots,T.
	\end{equation}
	
	These matrices are then double-centered to obtain
	\begin{equation}
		A_{mn} = a_{mn} - \bar{a}_{m\cdot} - \bar{a}_{\cdot n} + \bar{a}_{\cdot\cdot},
	\end{equation}
	\begin{equation}
		B_{mn} = b_{mn} - \bar{b}_{m\cdot} - \bar{b}_{\cdot n} + \bar{b}_{\cdot\cdot},
	\end{equation}
	where \(\bar{a}_{m\cdot}\) and \(\bar{b}_{m\cdot}\) denote the mean of row \(m\), \(\bar{a}_{\cdot n}\) and \(\bar{b}_{\cdot n}\) denote the mean of column \(n\), and \(\bar{a}_{\cdot\cdot}\) and \(\bar{b}_{\cdot\cdot}\) denote the grand means of the corresponding distance matrices.
	
	The sample distance covariance and distance variances are then defined as
	\begin{equation}
		\mathrm{dCov}^2(u,v) = \frac{1}{T^2} \sum_{m=1}^{T} \sum_{n=1}^{T} A_{mn} B_{mn},
	\end{equation}
	\begin{equation}
		\mathrm{dVar}^2(u) = \frac{1}{T^2} \sum_{m=1}^{T} \sum_{n=1}^{T} A_{mn}^2,
	\end{equation}
	\begin{equation}
		\mathrm{dVar}^2(v) = \frac{1}{T^2} \sum_{m=1}^{T} \sum_{n=1}^{T} B_{mn}^2.
	\end{equation}
	
	Distance correlation is then given by
	\begin{equation}
		\mathrm{DistCorr}(u,v) =
		\frac{\mathrm{dCov}(u,v)}
		{\sqrt{\mathrm{dVar}(u)\,\mathrm{dVar}(v)}},
	\end{equation}
	with the convention that \(\mathrm{DistCorr}(u,v)=0\) when the denominator is zero (i.e. statistical independence).
	
	In this formulation, larger DistCorr values indicate stronger statistical dependence between the original and reconstructed signals. We used this metric because it provides a flexible measure of reconstruction quality that is sensitive to dependence structure beyond purely linear correspondence.
	
	\subsection{Surrogate Data: Phase Shuffle, IAAFT, and Block-Shuffle}
	To further validate that the model leverages meaningful temporal and cross-channel structure, we generated surrogate datasets using three techniques: phase shuffle \cite{Theiler1992, Paeske2018}, Iterative Amplitude-Adjusted Fourier Transform (IAAFT) \cite{SchreiberSchmitz1996, Keylock2006}, and block-shuffle \cite{Kunsch1989, Nakamura2006}. These surrogates serve as controls by preserving certain statistical properties of the original signals while selectively disrupting temporal dynamics or cross-channel dependencies \cite{Theiler1992, SchreiberSchmitz1996}. Algorithm~\ref{alg_surrogate} shows the detailed implementation for generating each type of surrogate data.
	
	\paragraph{Phase-Shuffled Surrogate:}
	This method preserves the amplitude spectrum of each electrode while removing phase information, disrupting the temporal dynamics and oscillatory structure. Surrogate signals were generated by applying a Fourier transform to the original time series, permuting the phase components at positive frequencies, and maintaining Hermitian symmetry to ensure a real-valued output \cite{Theiler1992}. To mitigate artifacts in cyclic biological signals, we utilized end-matched segmentation as recommended by \cite{Paeske2018}. This approach ensures that model performance on surrogate data is not attributed to original temporal dependencies, but rather to residual statistical structures like power spectra.
	
	\paragraph{IAAFT:}
	To generate surrogate signals that preserve key statistical properties while disrupting temporal and cross-channel structure, we applied the Iterative Amplitude-Adjusted Fourier Transform (IAAFT) procedure \cite{SchreiberSchmitz1996}. Unlike standard phase randomization, IAAFT alternates a fixed number of times (\(n_{\mathrm{iter}} = 5\) in our 
	implementation) between satisfying the target power spectrum and the exact 
	amplitude distribution (histogram) of the original signal \cite{Keylock2006}. 
	This produces surrogates that retain the precise marginal distribution of each 
	electrode and approximately preserve its autocorrelation, while effectively 
	randomizing phase information and destroying meaningful nonlinear temporal 
	dependencies.
	
	\paragraph{Block-Shuffle:}
	Block-shuffle surrogates preserve short-term, local trends and internal dynamics within blocks while destroying long-range temporal order \cite{Kunsch1989}. Each electrode’s time series is divided into non-overlapping blocks of length $B$ seconds, which are then randomly permuted. This method provides a null hypothesis that tests whether the model relies on the specific chronological sequence of events or whether the information contained within isolated local windows is sufficient for reconstruction \cite{Nakamura2006, Small2001}.
	
	\vspace{4mm}
	
	\stepcounter{myalg}
	\makeatletter\def\@currentlabel{\themyalg}\makeatother\label{alg_surrogate} 
	\noindent\hrule
	\vspace{1mm}
	\noindent \textbf{Algorithm \themyalg:} Generation of phase-shuffled, IAAFT, and block-shuffled surrogate data
	\vspace{1mm}
	\hrule
	\vspace{1mm}
	\begin{algorithmic}[1]
		\Require Dataset $X \in \mathbb{R}^{\mathrm{trials} \times \mathrm{channels} \times \mathrm{samples}}$, surrogate method $\in \{\text{Phase}, \text{IAAFT}, \text{Block}\}$, block size $B$ (seconds), sampling frequency $f_s$, maximum number of IAAFT iterations $n_{\mathrm{iter}}$
		\Ensure Surrogate dataset $X_{\mathrm{surrogate}}$
		\Statex \hrulefill
		
		\For{each trial $t$ in $X$}
		\For{each channel $c$ in trial $t$}
		\State $S \gets X[t,c,:]$
		
		\If{\textbf{method = Phase}}
		\State $\mathcal{F}_S \gets \mathrm{FFT}(S)$
		\State $A \gets |\mathcal{F}_S|$
		\State $\phi \gets \angle \mathcal{F}_S$
		\State Permute the positive-frequency phases in $\phi$ 
		\State \hspace{1em} (or, more generally, replace them with randomized phases according to the chosen phase-shuffling scheme)
		\State Enforce Hermitian symmetry; preserve the DC component (and Nyquist frequency when applicable)
		\State $\mathcal{F}_{\mathrm{surr}} \gets A \cdot e^{i\phi}$
		\State $S_{\mathrm{surrogate}} \gets \mathrm{Re}\!\left(\mathrm{IFFT}(\mathcal{F}_{\mathrm{surr}})\right)$
		
		\ElsIf{\textbf{method = IAAFT}}
		\State $S_{\mathrm{sorted}} \gets \mathrm{sort}(S)$
		\State $A \gets |\mathrm{FFT}(S)|$
		\State $z \gets$ random permutation of $S$
		\For{$k = 1$ to $n_{\mathrm{iter}}$} 
		\State $Z \gets \mathrm{FFT}(z)$
		\State $Z \gets A \cdot e^{i\angle Z}$
		\State $z_{\mathrm{new}} \gets \mathrm{Re}\!\left(\mathrm{IFFT}(Z)\right)$
		\State Reorder $S_{\mathrm{sorted}}$ to match the rank ordering of $z_{\mathrm{new}}$
		\State $z \gets$ reordered signal
		\EndFor
		\State $S_{\mathrm{surrogate}} \gets z$
		
		\ElsIf{\textbf{method = Block}}
		\State $B_{\mathrm{samples}} \gets B \cdot f_s$
		\State Divide $S$ into contiguous blocks of length $B_{\mathrm{samples}}$
		\State Randomly permute the block order
		\State Concatenate the permuted blocks
		\State Append any remaining samples at the end, if present
		\State $S_{\mathrm{surrogate}} \gets$ resulting signal
		\EndIf
		
		\State $X_{\mathrm{surrogate}}[t,c,:] \gets S_{\mathrm{surrogate}}$
		\EndFor
		\EndFor
		
		\State \Return $X_{\mathrm{surrogate}}$
	\end{algorithmic}
	\hrule
	\vspace{2mm}

	\section{Results}
	\begin{figure*}[t]
		\begin{center}
			\includegraphics[width=\textwidth]{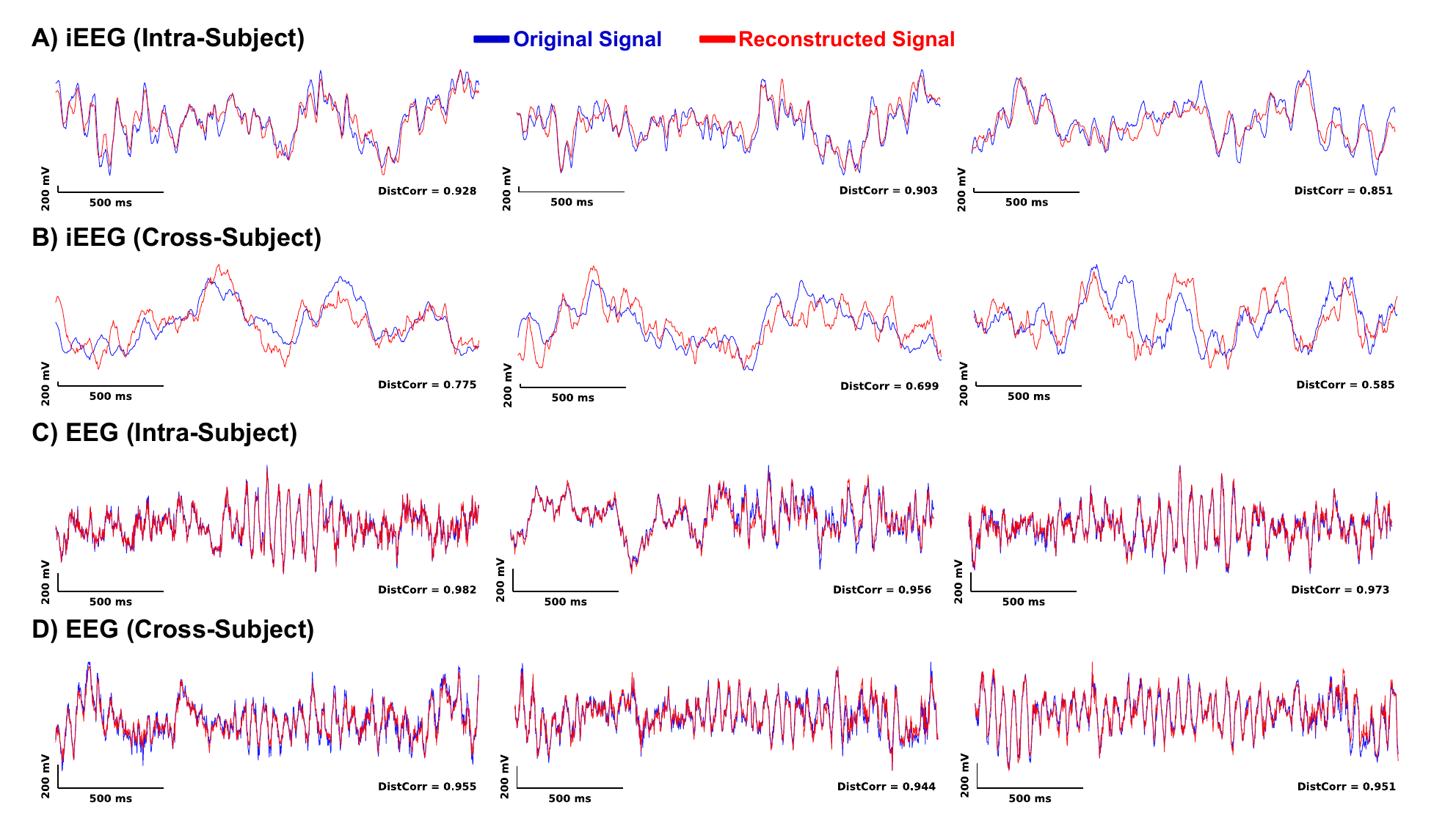}
		\end{center}
		\caption{Representative original (blue) and reconstructed (red) signals for a single channel under four conditions: (A) iEEG, intra-subject; (B) iEEG, cross-subject; (C) EEG, intra-subject; and (D) EEG, cross-subject. These examples illustrate SMR's reconstruction quality across recording modalities and transfer settings.}
		\label{fig3}
	\end{figure*}
	
	\subsection{Performance of the SMR Model in Intra-Subject Setting}
	We first evaluated SMR in the intra-subject setting, where each subject was modeled separately and performance was assessed on held-out data from the same individual. Mean DistCorr across subjects was \(0.908 \pm 0.028\) for the EEG dataset and \(0.553 \pm 0.068\) for the iEEG dataset. Subject-wise results are summarized in Fig.~\ref{fig2}, and representative examples of reconstructed versus original signals are shown in Fig.~\ref{fig3}A,C.
	
	The higher reconstruction performance in EEG likely reflects the greater spatial smoothness and channel-to-channel redundancy of scalp recordings, which are shaped by volume conduction through the head tissues; as a result, multiple scalp electrodes can share overlapping information from common neural sources \cite{Buzsaki2012extracellular,BretteDestexhe2012}.
	
	\begin{figure*}[t]
		\begin{center}
			\includegraphics[width= \linewidth]{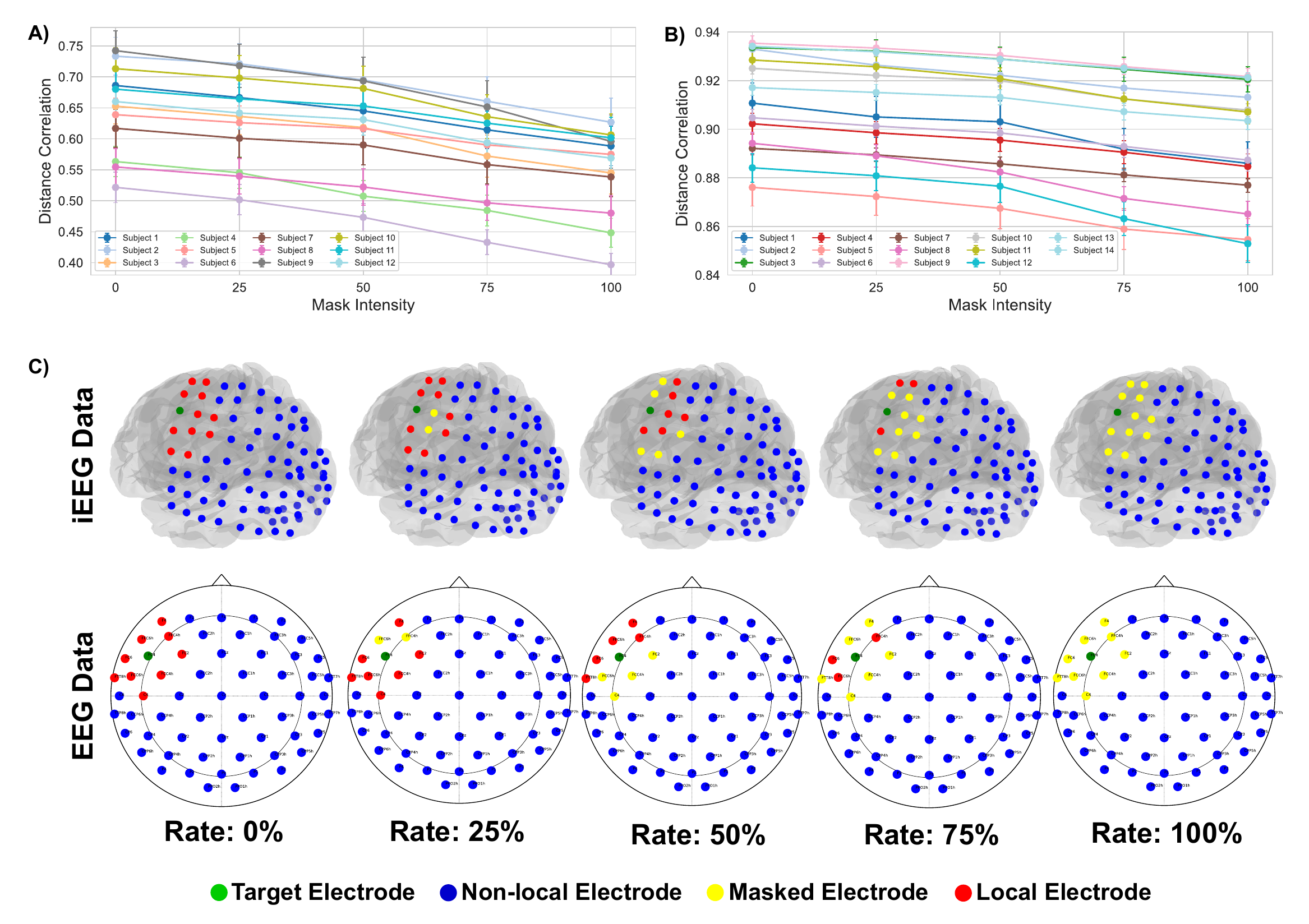}
		\end{center}
		\caption{Effect of local masking on reconstruction performance for (A) iEEG and (B) EEG. Distance correlation (DistCorr) is shown for mask intensities of 0\%, 25\%, 50\%, 75\%, and 100\%. Performance decreases as progressively more electrodes from the target channel’s predefined local neighborhood are masked, indicating that nearby channels carry substantial predictive information. Insets show representative masking schemes: blue, all electrodes; green, target electrode; red, the target’s local neighborhood; yellow, the subset masked at the indicated intensity.}
		\label{fig4}
	\end{figure*}

	\subsection{Performance of the SMR Model in Cross-Subject Mode}
	We next evaluated whether an inter-electrode relationship matrix learned from one subject could be successfully transferred to another without fitting target-specific weights. As expected, performance decreased relative to the intra-subject setting, but the model retained appreciable predictive power, with the magnitude of the transfer penalty differing markedly between EEG and iEEG.
	
	For the EEG dataset, we randomly selected a reference subject \(s'\) for each 
	target subject \(s\) and transferred the learned matrix directly, using the 
	reference subject's checkpoint trained at mask intensity \(m = 1.0\) (full 
	local-neighborhood masking). Because EEG electrode positions are standardized, 
	this provided a natural test of the model's ability to generalize under a common 
	sensor layout. The model achieved an average cross-subject DistCorr of 
	\(0.783 \pm 0.093\), indicating robust transfer of the learned inter-electrode 
	structure.
	
	For the iEEG dataset, where electrode coverage is heterogeneous and clinically determined, direct transfer is infeasible. To address this, we used the CBEM procedure (Section~\ref{cross_sub_eval}) to construct candidate transferred matrices from all available reference subjects. We then selected the most suitable matrix using a held-out validation subset from the target subject. Crucially, this step involved only selection among existing transferred matrices and did not involve any weight optimization on the target data. In this setting, the model achieved an average DistCorr of \(0.389 \pm 0.064\).
	
	The contrast between these modalities is informative. In EEG, the standardized montage and the fact that scalp recordings reflect broader spatial mixtures of underlying neural activity favor the transfer of learned inter-electrode structure across individuals. In iEEG, by contrast, the recordings are more spatially focal and subject-specific. Furthermore, the inherent variability in clinical electrode placement makes establishing exact functional correspondence across subjects fundamentally more difficult. These factors likely account for the lower cross-subject performance observed in iEEG. Representative examples of reconstructed signals are shown in Fig.~\ref{fig3}B,D, and detailed subject-wise averages are provided in Appendix Tables~\ref{tab:ieeg} and~\ref{tab:eeg}.
	
	\subsection{Performance of Lagged Masked Regression}
	We also evaluated a lagged variation of the SMR framework to test whether incorporating temporal offsets would improve reconstruction by capturing short-timescale dependencies and delayed interactions between channels. Because the analyzed epochs were motor related, the lag values were chosen to sample time scales relevant to \(\beta\)- and \(\gamma\)-range activity in motor cortex: \(\beta\) (15--30~Hz) and \(\gamma\) (30--110~Hz) \cite{Babiloni2016rhythm}. Specifically, we used \(L = [0, 20, 30, 50, 60]\)~ms, spanning delays broadly matched to the cycle durations of these rhythms. The corresponding lag in samples was computed as
		$\text{lag}_{i} = \frac{L(i)}{1000} f_s,$
	where \(f_s\) denotes the sampling frequency.
	
	As shown in Tables~\ref{tab:ieeg} and~\ref{tab:eeg}, the lagged model produced only minimal changes relative to the instantaneous baseline. In the intra-subject setting, EEG performance was \(0.910 \pm 0.027\), compared with \(0.908 \pm 0.028\) for the instantaneous model. For iEEG, the corresponding values were \(0.554 \pm 0.067\) and \(0.553 \pm 0.068\), respectively.
	
	Thus, within the lag range tested here, adding delayed predictors did not materially improve reconstruction. This suggests that most of the recoverable structure for this task was already captured by the instantaneous model, or that any additional lagged contributions were too weak, too redundant, or too transient to alter the summary performance measure appreciably. We therefore view the lagged analysis as a secondary extension of the framework rather than a primary source of performance gain. Even so, it remains useful as a way to probe whether short-timescale temporal offsets contribute additional predictive structure beyond the instantaneous relationships captured by the baseline SMR model.
	
	\subsection{Analysis of the Effect of Local Information}
	
	We next examined how reconstruction depends on the predefined local neighborhood of each target electrode. Rather than masking arbitrary channels, this analysis progressively excluded electrodes belonging to the local neighborhood associated with the target, as defined in Section~\ref{model_arch}. In this way, mask intensity served as a controlled probe of how strongly reconstruction relied on spatially local information.
	
	We tested five mask intensities \(m \in \{0,\,0.25,\,0.50,\,0.75,\,1.0\}\). 
	For each target electrode \(i\), 
	\(\lceil m \cdot |N^{(s)}(i)| \rceil\) electrodes were drawn uniformly at random 
	from the local neighborhood \(N^{(s)}(i)\) and masked together with \(i\) itself; 
	the remaining channels served as predictors. Thus \(m=0\) corresponds to keeping 
	all neighborhood channels available (only the target is masked), and \(m=1.0\) 
	corresponds to masking the entire neighborhood. Intermediate values represent 
	progressively stronger removal of local information.
	
	The results for both datasets are shown in Fig.~\ref{fig4}. In each case, DistCorr decreased as mask intensity increased. This pattern indicates that reconstruction depends strongly on the local neighborhood of the target electrode: when progressively more of the spatially related nearby channels are removed, performance declines accordingly. Thus, the model is not relying on arbitrary global correlations alone, but makes substantial use of the structured local neighborhood assigned to each target electrode. At the same time, reconstruction remains above zero even under full local masking, indicating that the target signal is not determined solely by its immediate neighborhood. Thus, the results support a mixed local--distributed picture: locally structured information contributes strongly to reconstruction, but measurable predictive structure also remains available outside the masked neighborhood.
	
	\subsection{Comparison of Electrode Coverage}
	\label{sec:electrode_comparison}
	We compared model performance across three input configurations: (i) \textit{local electrodes}, where only spatially adjacent electrodes were used, (ii) \textit{non-local electrodes}, where distant electrodes were included but local neighbors were masked, and (iii) \textit{all electrodes}, where both local and non-local inputs were available. 
	
	\begin{figure}[t]
		\centering
		\includegraphics[width=\columnwidth]{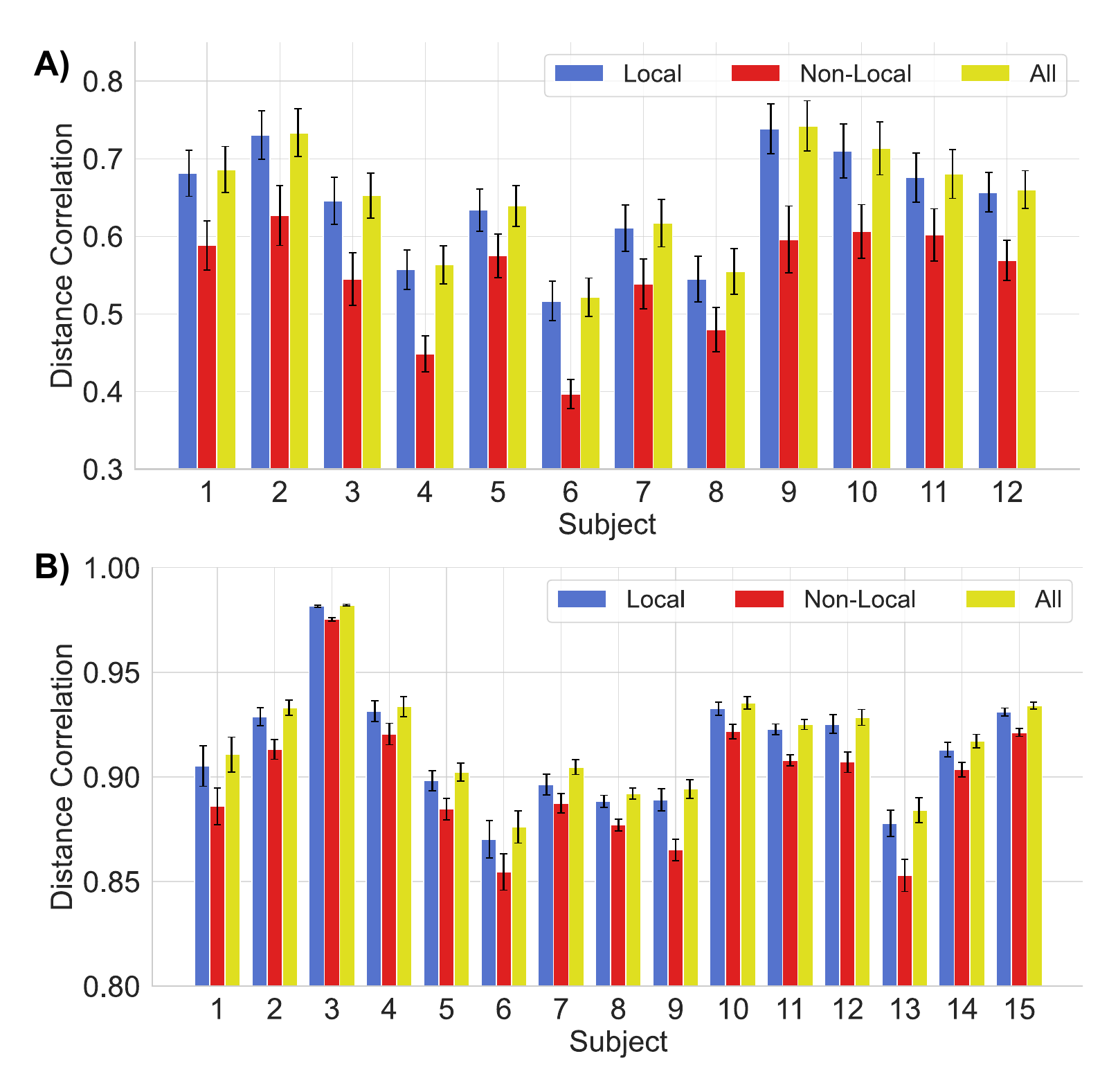}
		\caption{Comparison of reconstruction performance across three input configurations for (A) iEEG and (B) EEG. \textit{Local} uses only spatially adjacent electrodes, \textit{Non-local} excludes local neighbors and uses distant electrodes, and \textit{All} includes the full electrode set.}
		\label{fig6}
	\end{figure}
	
	Figure~\ref{fig6} presents model performance across these configurations. As expected, the best performance was achieved when both local and non-local inputs were available, indicating that short-range and long-range spatial features provide complementary predictive structure. Notably, the model trained with only local electrodes consistently outperformed the non-local configuration. This suggests that the immediate spatial neighborhood contains more concentrated and coherent predictive information than distant sensors, likely due to the preservation of fine-grained spatial structure. However, the superior performance of the all-electrode condition demonstrates that non-local information still contributes meaningfully when integrated with local inputs, capturing broader spatial dynamics that local sensors alone cannot resolve.
	
	\subsection{Surrogate Data Analysis}
	To determine whether the SMR model's performance relies on structured temporal organization rather than simple marginal signal statistics, we repeated the intra-subject analysis on three classes of surrogate datasets: phase-shuffled, Iterative Amplitude Adjusted Fourier Transform (IAAFT), and block-shuffled. These controls are designed to preserve specific low-order properties of the original data while systematically disrupting phase relationships, temporal ordering, or both.
	
	Specifically, \textbf{phase-shuffled} surrogates preserve the power spectrum (and thus the linear autocorrelation) while destroying phase information and fine-grained temporal dynamics. \textbf{IAAFT} surrogates provide a more stringent control by preserving both the marginal amplitude distribution and the approximate power spectrum, effectively isolating the contribution of non-linear temporal dependencies. Finally, \textbf{block-shuffled} surrogates retain short-term local trends while disrupting long-range temporal order and cross-channel coordination.
	
	The model was evaluated on each surrogate dataset using the same training and evaluation pipeline as for the original neural recordings, in an intra-subject setting. Across all surrogate conditions, performance dropped substantially relative to the original recordings. For iEEG (Table~\ref{tab:ieeg}), the mean DistCorr decreased from \(0.553 \pm 0.068\) for the original data to \(0.181 \pm 0.029\), \(0.232 \pm 0.060\), and \(0.201 \pm 0.168\) for phase-shuffled, IAAFT, and block-shuffled surrogates, respectively. For EEG (Table~\ref{tab:eeg}), the corresponding decrease was from \(0.908 \pm 0.028\) to \(0.124 \pm 0.049\), \(0.184 \pm 0.073\), and \(0.160 \pm 0.078\).
	
	These results collectively show that SMR performance is not merely a reflection of marginal signal statistics, such as amplitude distributions or static power spectra. The strongest average degradation was observed under phase shuffling, indicating that temporal phase structure contributes importantly to reconstruction. IAAFT surrogates preserved more low-order properties and therefore retained somewhat more reconstructability, but performance still remained substantially below that of the original data. Block shuffling also reduced performance markedly, indicating that longer-range temporal ordering contributes beyond short-term local trends. Taken together, these controls support the view that SMR is not driven by amplitude distributions or power spectra alone; its performance depends on structured temporal and cross-channel organization present in the original recordings.
	
	\section{Discussion}
	
	Spatially extended recordings sit at a conceptual fault line: every channel looks ``local'' when interpreted as a field potential measured at a specific location, yet the same channel often participates in rhythms and patterns that are distributed across large-scale networks \cite{Buzsaki2004Oscillations,Bullmore2009Complex}. SMR makes this tension measurable by asking a direct question about redundancy and representation: how much of a channel can be reconstructed from the rest, and how does that reconstructability decay as local information is removed? In doing so, the framework turns a common nuisance in multichannel recordings---strong short-range similarity---into a probe of broader distributed structure.
	
	A first implication of the present results is conceptual. The masking analysis provides a concrete handle on the local--global continuum in field recordings. Removing nearby electrodes systematically degraded reconstruction, confirming that local neighborhoods carry substantial predictive information, as expected from the smoothness of cortical fields \cite{Einevoll2024electric} and the spatial mixing inherent in EEG and many iEEG configurations \cite{Srinivasan2007ScalesCoherence,Nolte2004ImagCoherency}. At the same time, the persistence of nontrivial reconstruction performance under strict local exclusion argues against an interpretation of channels as isolated readouts of purely local generators. Instead, EEG and iEEG signals behave like projections of a distributed dynamical system whose state is redundantly represented across space \cite{Buzsaki2004Oscillations,Bullmore2009Complex}. In electrophysiology, where local correlations are often dismissed as volume-conduction artifacts, SMR helps separate two distinct facts: (i) local redundancy is expected from biophysics, and (ii) beyond that redundancy, distributed structure can still carry predictive content \cite{Schoffelen2009SourceConnectivity,Nolte2004ImagCoherency}.
	
	The comparison across electrode-coverage conditions further sharpens this point. Local electrodes alone consistently outperformed non-local electrodes alone, indicating that the most concentrated predictive information is carried by the immediate spatial neighborhood of the target channel. At the same time, the best performance was obtained when both local and non-local inputs were available, showing that distant electrodes contribute additional predictive structure once the local context is present. This result argues against two overly simple views: first, that reconstruction is driven purely by broad distributed correlations, and second, that the target signal is determined entirely by its nearest neighbors. Instead, the findings support a hierarchical local--distributed organization in which nearby electrodes provide the strongest core information, while more distant channels add complementary large-scale structure. From a practical standpoint, this suggests that reconstruction and decoding strategies may benefit from preserving dense local coverage while also retaining enough broader sampling to capture distributed network-level structure.
	
	This distinction is sharpened when viewed through the lens of field biophysics. Extracellular filtering and tissue properties impose frequency- and distance-dependent structure on measured potentials \cite{Bedard2004FreqFiltering,Bedard2009MacroscopicLFP1f}, and this helps explain why EEG and iEEG differ in reconstructability and transferability. Scalp EEG is shaped not only by broader volume conduction through the head tissues, which increases spatial overlap across channels \cite{Buzsaki2012extracellular,BretteDestexhe2012}, but also by frequency-dependent filtering along the path from generators to sensors. Theoretical work has shown that non-homogeneous conductivity and permittivity in extracellular media can produce low-pass filtering and distance-dependent attenuation \cite{Bedard2004FreqFiltering}, while measurements and parametric models of biological tissues have documented strong frequency dependence in their dielectric properties \cite{Gabriel1996dielectricI,Gabriel1996dielectricII,Gabriel1996dielectricIII}. Consistent with this view, simultaneous EEG--MEG analyses have shown differences in low-frequency spectral scaling that are difficult to reconcile with a purely resistive medium and are compatible with filtering contributions from the extracellular medium, including dura and skull \cite{Dehghani2010EEGMEGScaling}. Within this framework, scalp EEG can be understood as both more spatially mixed and more spectrally filtered than iEEG, making it unsurprising that short-range predictors are especially powerful and that inter-electrode relationships generalize more readily across subjects when the montage is standardized \cite{Jurcak2007EEGMontage}.
	
	The comparison between EEG and iEEG therefore further clarifies why ``local field'' is an incomplete descriptor. In the intra-subject setting, reconstruction was consistently stronger in EEG than in iEEG, and the cross-subject analysis reinforced this contrast: transfer remained substantial in EEG under a standardized montage, but was markedly weaker in iEEG. This cross-subject difference likely reflects two related factors. First, scalp EEG is shaped by broader volume conduction through the head tissues, which increases spatial overlap across channels and makes inter-electrode relationships more redundant \cite{Buzsaki2012extracellular,BretteDestexhe2012}. Second, EEG is typically recorded with a standardized montage, so homologous channels sample broadly comparable regions across individuals \cite{Jurcak2007EEGMontage}.
	
	By contrast, iEEG electrodes sample activity closer to the underlying generators and therefore retain more focal and subject-specific structure, reducing channel-to-channel redundancy. In addition, iEEG coverage varies substantially across subjects in both electrode placement and anatomical extent, so the recorded channel set itself is not held constant across individuals \cite{Peterson2022AJILE12}. This heterogeneity makes cross-subject transfer especially difficult, because learned inter-electrode relationships are less portable when both the signal structure and the sampled spatial geometry differ from one subject to another. This interpretation is also broadly consistent with earlier EEG--MEG comparisons suggesting that scalp EEG can exhibit a more synchronous and spatially coherent signature than a more source-selective modality, even when the underlying generators are more heterogeneous or asynchronous \cite{Dehghani2011spindleHBM,Dehghani2010spindleJNP,Dehghani2010spindlePLoS}. More broadly, these findings support the idea that the balance between local and distributed contributions is modality dependent: EEG appears to occupy a more redundant, spatially mixed regime, whereas iEEG preserves a more local and individualized representation of neural activity.
	
	What is more informative, however, is what remains when strong local predictors are withheld: the extent to which activity at distant sites continues to constrain the target channel, beyond what would be expected from trivial spatial smoothing alone \cite{Srinivasan2007ScalesCoherence,Schoffelen2009SourceConnectivity}. In this sense, SMR provides an empirical complement to biophysical arguments: it quantifies the ``effective reach'' of predictive structure in the data, and how that reach differs by modality and geometry.
	
	The minimal gain obtained with the lagged model suggests that, within the lag range tested here, most of the reconstructable structure in these motor-related data was already captured by the instantaneous SMR model. One plausible explanation is that the dominant inter-electrode dependencies relevant for reconstruction are sufficiently strong at zero lag that adding short delays contributes little additional information. This is consistent with the fact that motor-related $\beta$- and $\gamma$-range activity is often transient and state dependent \cite{LVQ2016hfo, Barone2021betaSensorimotor,Muthukumaraswamy2010gammaMotor}, with event-related bursts and rapid changes in synchrony shaping the observed signal structure \cite{Babiloni2016rhythm,Gascoyne2021motor,West2023burst}. Under such conditions, delayed predictors may add only limited information beyond what is already reflected in the instantaneous spatial covariance pattern. At the same time, the absence of a large performance gain should not be taken to mean that short-timescale temporal interactions are unimportant. Rather, it suggests that the present reconstruction objective and summary metric are relatively insensitive to those additional contributions at the level of overall signal recoverability. In this sense, the lagged model is best viewed as a useful extension for probing whether delayed cross-channel dependencies contribute predictive structure beyond the baseline SMR formulation, rather than as a mechanism that necessarily improves reconstruction accuracy in a large or systematic way. We therefore do not interpret the resulting lagged weights ``causally''; they are more appropriately understood as reflecting time-offset statistical dependencies within the multichannel recordings.
	
	The surrogate analysis provides an important constraint on how the present model should be interpreted. The marked reduction in performance across all three surrogate classes indicates that SMR is not effective merely by exploiting static marginal properties of the recordings, such as amplitude distributions or coarse spectral content. Rather, successful reconstruction depends on temporal organization that is progressively disrupted when phase relationships are randomized, higher-order temporal dependencies are removed, or longer-range ordering is broken. The especially strong effect of phase shuffling suggests that temporally structured phase relationships contribute importantly to the predictive signal available to the model. At the same time, the fact that IAAFT and block-shuffled surrogates also remain well below the performance obtained on the original recordings indicates that preserving low-order statistics alone is insufficient. Taken together, these findings support the view that SMR exploits structured spatiotemporal dependencies present in the original data, rather than trivial signal regularities alone. The surrogate analysis further suggests that the structure exploited by SMR is not captured by stationary low-order signal properties alone. While these controls do not establish non-stationarity in a strict formal sense, the marked degradation under phase randomization and temporal reordering is consistent with the presence of time-varying temporal organization in the original recordings. This fits naturally with the transient and event-linked character of motor-related neural activity \cite{LVQ2016hfo,Barone2021betaSensorimotor,Muthukumaraswamy2010gammaMotor}.
	
	Methodologically, SMR complements functional connectivity analyses by shifting the primitive from pairwise association to predictive embedding. Traditional measures such as coherence or PLV quantify synchronization between pairs \cite{Srinivasan2007ScalesCoherence,Lachaux1999PLV}, while information-theoretic measures target broader statistical dependence and directed influence \cite{Ince2017GCMI,Vicente2011TransferEntropy}. Graph approaches then summarize these relations into network-level descriptors \cite{Bullmore2009Complex}, often with important interpretational caveats \cite{Hallquist2019BraveNewSmallWorld}. SMR asks a different question: given all other channels, how well is a target channel determined? This reconstructability viewpoint naturally yields an asymmetric weight matrix and admits spatial masking as an explicit control that suppresses short-range contributions. In that sense, SMR does not compete with functional connectivity so much as diagnose which aspects of apparent interdependence are driven by proximity and which reflect broader embedding in distributed dynamics.
	
	The reconstruction viewpoint also connects naturally to model-based inference of missing activity. Gaussian-process approaches infer activity at unobserved locations by learning spatial covariance structure \cite{Owen2020GaussianProcessECoG}; latent dynamical models such as LFADS exploit low-dimensional dynamics for denoising and reconstruction in neural population activity \cite{Pandarinath2018LFADS}; sparse autoregressive models seek directed interactions under high-dimensional constraints \cite{ValdesSosa2005SparseMVAR}. SMR sits deliberately at a simpler point in this spectrum: it is linear, it is explicit, and it is intentionally constrained by spatial masks. That simplicity is a feature for interpretability and for making local exclusion an experimental knob. At the same time, it points to immediate extensions: kernelized regression or nonlinear predictors could capture dependencies beyond linear structure, while preserving the same masking protocol for controlled removal of local information \cite{Roy2019DLEEGReview}.
	
	Several directions follow naturally. First, SMR can guide sensor selection and montage design by identifying channels whose activity is highly predictable from the rest, and therefore relatively redundant, versus channels that contribute more unique information. This is relevant both for practical constraints in BCIs and for experimental designs where dense coverage is infeasible. Second, the framework can be made state-aware. The balance between local and global dependencies is expected to vary across brain states and behavioral contexts; work on excitation--inhibition dynamics across states \cite{Peyrache2012EIDynamicsSleep,Dehghani2016EIBalance} motivates applying masked reconstruction as a quantitative descriptor of how distributed coordination reorganizes in sleep, wake, and pathology. Third, extending SMR to incorporate richer temporal structure---for example through state-space predictors or autoregressive embeddings---would allow the same spatial masking logic to interrogate not only instantaneous coupling but also lagged predictability, bringing the framework into closer contact with directed interaction models \cite{ValdesSosa2005SparseMVAR,Vicente2011TransferEntropy}.
	
	\subsection*{Conclusion}
	Taken together, the present results support a mixed local--distributed view of electrophysiological signals. In the intra-subject setting, reconstruction was consistently stronger in EEG than in iEEG, likely reflecting the greater spatial smoothness and channel-to-channel redundancy of scalp recordings, whereas iEEG remains closer to the underlying generators and therefore preserves more focal and subject-specific structure. The cross-subject analysis reinforced this contrast: transfer remained substantial in EEG under a standardized montage, but was markedly weaker in iEEG, where variable electrode coverage across subjects, together with greater spatial specificity, limits the portability of learned inter-electrode structure. The masking and electrode-coverage analyses further clarified this picture by showing that nearby electrodes carry the most concentrated predictive information, yet that optimal performance requires combining local and non-local inputs. Thus, electrode signals are not explained by purely local redundancy alone, nor by broad distributed correlations alone, but by a balance between the two. Finally, the surrogate analysis showed that SMR is not merely exploiting static marginal properties of the signals. Its performance depended strongly on structured temporal and cross-channel organization in the original recordings, with substantial degradation under phase shuffling, IAAFT, and block shuffling. Overall, these findings suggest that SMR provides an interpretable framework for quantifying how local neighborhood structure and broader distributed organization jointly shape electrophysiological recordings across modalities.
	
	\begin{acknowledgments}
		N.D. wishes to acknowledge the support of NIH grant R24MH117295.
	\end{acknowledgments}
	
	\section*{Data Availability}
	The iEEG data used in this study are publicly available as part of the AJILE12 dataset, which can be accessed through the DANDI (Distributed Archive for Neurophysiology Data Integration) repository at \href{https://dandiarchive.org/dandiset/000055/}{DANDISET~000055}. The EEG data used for comparison and supplementary analysis were obtained from the BNCI Horizon 2020 database: \href{https://bnci-horizon-2020.eu/database/data-sets}{BNCI\_Horizon\_2020}.
	
	\section*{Code Availability}
	The code used to preprocess the data, SMR model, and train the networks is available at GitHub \href{https://github.com/neurovium/SpatiallyMaskedRegression}{\texttt{https://github.com/neurovium/SpatiallyMaskedRegression}}. All scripts required to reproduce the analyses and results presented in this study are provided, along with documentation and example usage.
	
	
	\begin{table*}[h!]
		\centering
		\setlength{\tabcolsep}{5pt} 
		\renewcommand{\arraystretch}{1.5} 
		\caption{Subject-wise distance correlation (DistCorr; mean $\pm$ SD across electrodes) for iEEG in the intra-subject, lagged, cross-subject, and surrogate conditions. \textbf{Note:} the Intra-Subject, Lagged, Phase Shuffle, IAAFT, and Block Shuffle columns report the mean across single-trial DistCorr values; the Cross-Subject column reports DistCorr on trial-averaged signals (see §\ref{cross_sub_eval}).}
		\label{tab:ieeg}
		\begin{tabular}{@{}c c c c c c c@{}}
			\hline
			\textbf{Subject} & \textbf{Intra-Subject} & \textbf{Lagged} & \textbf{Cross-Subject} & \textbf{Phase Shuffle} & \textbf{IAAFT} & \textbf{Block Shuffle} \\
			\hline
			1 & 0.585 $\pm$ 0.041 & 0.590 $\pm$ 0.041 & 0.331 $\pm$ 0.007 & 0.176 $\pm$ 0.007 & 0.213 $\pm$ 0.005 & 0.148 $\pm$ 0.007 \\
			2 & 0.647 $\pm$ 0.038 & 0.649 $\pm$ 0.038 & 0.483 $\pm$ 0.013 & 0.252 $\pm$ 0.013 & 0.228 $\pm$ 0.005 & 0.112 $\pm$ 0.002 \\
			3 & 0.535 $\pm$ 0.035 & 0.559 $\pm$ 0.036 & 0.440 $\pm$ 0.012 & 0.207 $\pm$ 0.008 & 0.344 $\pm$ 0.039 & 0.698 $\pm$ 0.013 \\
			4 & 0.452 $\pm$ 0.036 & 0.451 $\pm$ 0.036 & 0.325 $\pm$ 0.006 & 0.176 $\pm$ 0.005 & 0.156 $\pm$ 0.009 & 0.397 $\pm$ 0.013 \\
			5 & 0.524 $\pm$ 0.034 & 0.515 $\pm$ 0.034 & 0.509 $\pm$ 0.016 & 0.149 $\pm$ 0.013 & 0.220 $\pm$ 0.003 & 0.129 $\pm$ 0.004 \\
			6 & 0.428 $\pm$ 0.021 & 0.428 $\pm$ 0.022 & 0.285 $\pm$ 0.004 & 0.149 $\pm$ 0.003 & 0.111 $\pm$ 0.003 & 0.126 $\pm$ 0.008 \\
			7 & 0.531 $\pm$ 0.028 & 0.533 $\pm$ 0.029 & 0.415 $\pm$ 0.010 & 0.187 $\pm$ 0.008 & 0.297 $\pm$ 0.002 & 0.159 $\pm$ 0.015 \\
			8 & 0.487 $\pm$ 0.025 & 0.489 $\pm$ 0.026 & 0.377 $\pm$ 0.013 & 0.184 $\pm$ 0.007 & 0.220 $\pm$ 0.003 & 0.180 $\pm$ 0.003 \\
			9 & 0.635 $\pm$ 0.041 & 0.622 $\pm$ 0.042 & 0.409 $\pm$ 0.010 & 0.132 $\pm$ 0.011 & 0.295 $\pm$ 0.003 & 0.130 $\pm$ 0.011 \\
			10 & 0.605 $\pm$ 0.040 & 0.601 $\pm$ 0.041 & 0.363 $\pm$ 0.007 & 0.181 $\pm$ 0.008 & 0.228 $\pm$ 0.004 & 0.113 $\pm$ 0.002 \\
			11 & 0.612 $\pm$ 0.032 & 0.614 $\pm$ 0.032 & 0.395 $\pm$ 0.014 & 0.190 $\pm$ 0.008 & 0.213 $\pm$ 0.003 & 0.112 $\pm$ 0.002 \\
			12 & 0.593 $\pm$ 0.032 & 0.596 $\pm$ 0.032 & 0.340 $\pm$ 0.007 & 0.187 $\pm$ 0.005 & 0.253 $\pm$ 0.003 & 0.107 $\pm$ 0.002 \\
			\hline
			\textbf{Average} & \textbf{0.553 $\pm$ 0.068} & \textbf{0.554 $\pm$ 0.067} & \textbf{0.389 $\pm$ 0.064} & \textbf{0.181 $\pm$ 0.029} & \textbf{0.232 $\pm$ 0.060} & \textbf{0.201 $\pm$ 0.168} \\
			\hline
		\end{tabular}
	\end{table*}

	\begin{table*}[h!]
		\centering
		\setlength{\tabcolsep}{5pt} 
		\renewcommand{\arraystretch}{1.5} 
		\caption{Subject-wise distance correlation (DistCorr; mean $\pm$ SD across electrodes) for EEG in the intra-subject, lagged, cross-subject, and surrogate conditions.}
		\label{tab:eeg}
		\begin{tabular}{@{}c c c c c c c@{}}
			\hline
			\textbf{Subject} & \textbf{Intra-Subject} & \textbf{Lagged} & \textbf{Cross-Subject} & \textbf{Phase Shuffle} & \textbf{IAAFT} & \textbf{Block Shuffle} \\
			\hline
			1 & 0.897 $\pm$ 0.009 & 0.901 $\pm$ 0.008 & 0.784 $\pm$ 0.040 & 0.107 $\pm$ 0.004 & 0.213 $\pm$ 0.004 & 0.147 $\pm$ 0.007 \\
			2 & 0.925 $\pm$ 0.004 & 0.926 $\pm$ 0.004 & 0.782 $\pm$ 0.037 & 0.123 $\pm$ 0.004 & 0.130 $\pm$ 0.005 & 0.111 $\pm$ 0.002 \\
			3 & 0.980 $\pm$ 0.001 & 0.981 $\pm$ 0.000 & 0.522 $\pm$ 0.054 & 0.298 $\pm$ 0.032 & 0.341 $\pm$ 0.038 & 0.399 $\pm$ 0.013 \\
			4 & 0.927 $\pm$ 0.005 & 0.928 $\pm$ 0.004 & 0.635 $\pm$ 0.059 & 0.154 $\pm$ 0.008 & 0.352 $\pm$ 0.008 & 0.296 $\pm$ 0.012 \\
			5 & 0.893 $\pm$ 0.005 & 0.896 $\pm$ 0.004 & 0.770 $\pm$ 0.035 & 0.116 $\pm$ 0.003 & 0.123 $\pm$ 0.003 & 0.133 $\pm$ 0.005 \\
			6 & 0.864 $\pm$ 0.008 & 0.869 $\pm$ 0.007 & 0.783 $\pm$ 0.018 & 0.108 $\pm$ 0.003 & 0.112 $\pm$ 0.003 & 0.124 $\pm$ 0.008 \\
			7 & 0.894 $\pm$ 0.004 & 0.897 $\pm$ 0.004 & 0.823 $\pm$ 0.017 & 0.103 $\pm$ 0.005 & 0.204 $\pm$ 0.004 & 0.159 $\pm$ 0.015 \\
			8 & 0.883 $\pm$ 0.003 & 0.884 $\pm$ 0.003 & 0.837 $\pm$ 0.006 & 0.112 $\pm$ 0.003 & 0.116 $\pm$ 0.003 & 0.119 $\pm$ 0.003 \\
			9 & 0.882 $\pm$ 0.005 & 0.885 $\pm$ 0.004 & 0.773 $\pm$ 0.019 & 0.088 $\pm$ 0.002 & 0.194 $\pm$ 0.003 & 0.130 $\pm$ 0.011 \\
			10 & 0.925 $\pm$ 0.003 & 0.926 $\pm$ 0.003 & 0.895 $\pm$ 0.005 & 0.125 $\pm$ 0.004 & 0.131 $\pm$ 0.005 & 0.112 $\pm$ 0.002 \\
			11 & 0.918 $\pm$ 0.002 & 0.919 $\pm$ 0.002 & 0.899 $\pm$ 0.005 & 0.107 $\pm$ 0.002 & 0.113 $\pm$ 0.003 & 0.112 $\pm$ 0.002 \\
			12 & 0.918 $\pm$ 0.004 & 0.920 $\pm$ 0.004 & 0.831 $\pm$ 0.019 & 0.107 $\pm$ 0.003 & 0.154 $\pm$ 0.003 & 0.106 $\pm$ 0.002 \\
			13 & 0.872 $\pm$ 0.006 & 0.873 $\pm$ 0.006 & 0.750 $\pm$ 0.027 & 0.109 $\pm$ 0.004 & 0.179 $\pm$ 0.005 & 0.146 $\pm$ 0.013 \\
			14 & 0.909 $\pm$ 0.003 & 0.911 $\pm$ 0.003 & 0.840 $\pm$ 0.016 & 0.100 $\pm$ 0.002 & 0.202 $\pm$ 0.002 & 0.175 $\pm$ 0.010 \\
			15 & 0.928 $\pm$ 0.002 & 0.929 $\pm$ 0.002 & 0.819 $\pm$ 0.023 & 0.105 $\pm$ 0.003 & 0.189 $\pm$ 0.003 & 0.130 $\pm$ 0.009 \\
			\hline
			\textbf{Average} & \textbf{0.908 $\pm$ 0.028} & \textbf{0.910 $\pm$ 0.027} & \textbf{0.783 $\pm$ 0.093} & \textbf{0.124 $\pm$ 0.049} & \textbf{0.184 $\pm$ 0.073} & \textbf{0.160 $\pm$ 0.078} \\
			\hline
		\end{tabular}
	\end{table*}

	\begin{figure*}[h!]
		\centering
		\includegraphics[width=18 cm, height= 22cm]{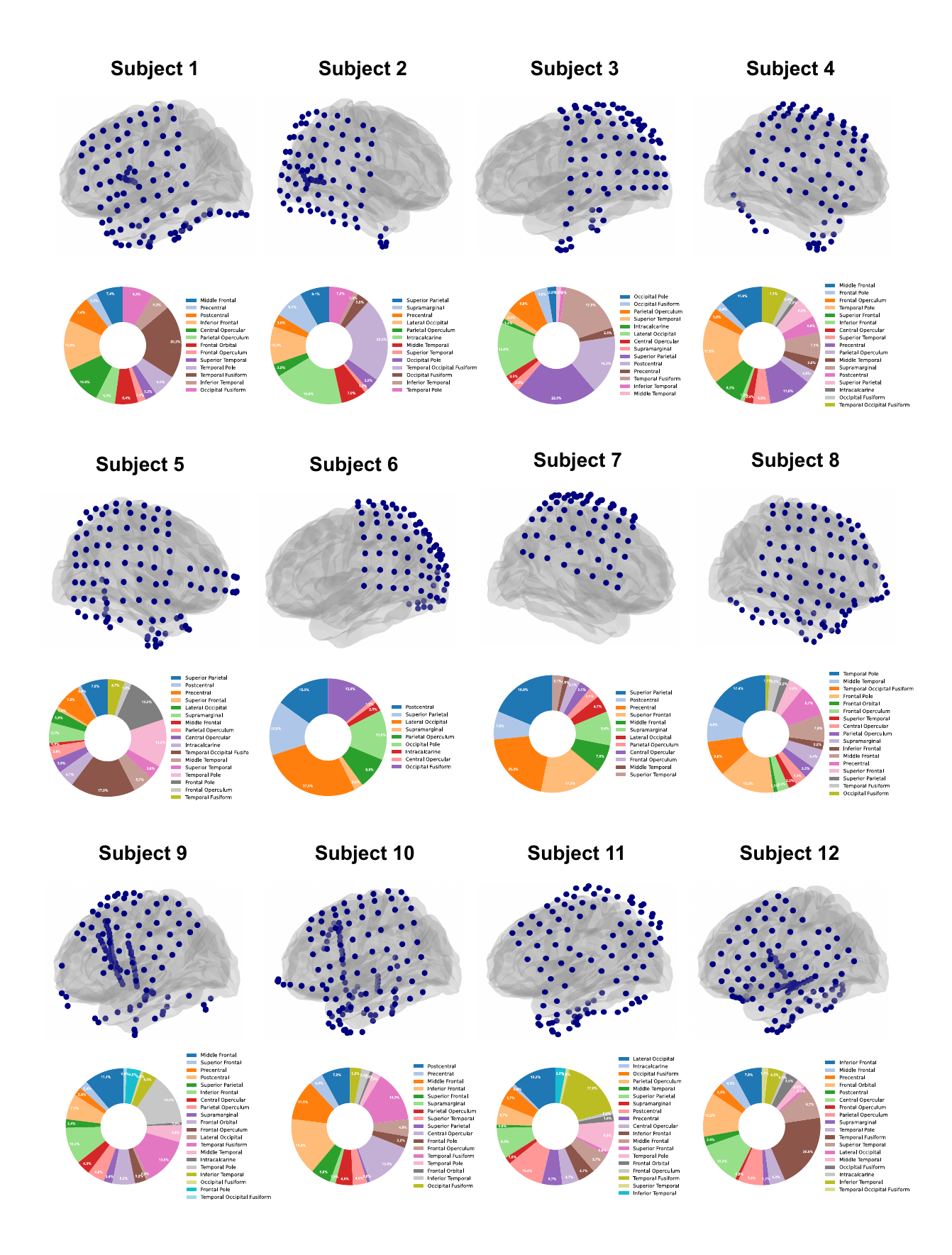} 
		\caption{\textbf{Electrode localization and regional distribution across subjects.} Three-dimensional cortical renderings show the spatial positions of intracranial EEG electrodes (blue dots) for each of the 12 subjects in AJILE dataset. The pie chart beneath each brain summarizes the proportion of electrodes assigned to major cortical regions based on anatomical labels (AAL). Together, these panels illustrate the substantial inter-subject variability in iEEG coverage across frontal, parietal, temporal, and occipital cortex.}
		\label{fig_app}
	\end{figure*}

	\section*{References}
	\bibliography{SMR_local_global}
	
\end{document}